\documentclass[final,3p,twocolumn,times]{elsarticle}

\usepackage{amssymb}
 \usepackage{amsthm}
\usepackage{cancel}

\usepackage{hyperref}
\usepackage{verbatim}

\usepackage{mathrsfs}
\usepackage{amsmath}
\usepackage{bm}
\usepackage{graphicx}
\usepackage{color}
\usepackage{multirow}

 \biboptions{square,sort&compress,comma}

\newcounter{bla}

\journal{Computer Physics Communications}

\begin{document}

\begin{frontmatter}



\title{BORAY: An Axisymmetric Ray Tracing Code Supports Both Closed and Open Field Lines Plasmas}




\author[label1,label2]{Hua-sheng XIE}
\ead{huashengxie@gmail.com, xiehuasheng@enn.cn}

\author[label1,label2]{Banerjee Debabrata}

\author[label1,label2]{Yu-kun BAI}

\author[label1,label2]{Han-yue ZHAO}

\author[label3]{Jing-chun LI}
 
\address[label1]{Hebei Key Laboratory of Compact Fusion, Langfang 065001, China}
\address[label2]{ENN Science and Technology Development Co., Ltd., Langfang 065001, China}
\address[label3]{Department of Earth and Space Sciences, Southern University of Science and Technology, Shenzhen, China}

\begin{abstract}
Ray tracing codes are useful to study the electromagnetic wave propagation and absorption in the geometrical optics approximation. In magnetized fusion plasma community, most ray tracing codes assume the plasma density and temperature be functions of the magnetic flux and study waves only inside the last closed flux surface, which are sufficient for the present day tokamak. However, they are difficult to be used for configurations with open magnetic field line plasmas, such as mirror machine and field-reversed-configuration (FRC). We develop a ray tracing code in cylindrical coordinates $(r,\phi,z)$ to support arbitrary axisymmetric configurations with both closed and open field lines plasmas. For wave propagation, the cold plasma dispersion relation is usually sufficient, and we require the magnetic field ${\bm B}(r,z)$ and species densities $n_{s0}(r,z)$ profiles as input. For wave absorption, we require a further temperature  $T_{s0}(r,z)$ profile to solve a hot kinetic plasma dispersion relation. In difference to other ray tracing codes which calculate the imaginary part of wave vector ${\bm k}_{\perp,i}$ for wave absorption, we calculate the imaginary part of wave frequency $\omega_i$, which are shown to be equivalent with the former technique under weak damping approximation. The code can use either numerical or analytical equilibrium. Examples and benchmarks with electron cyclotron wave, lower hybrid wave and ion cyclotron wave for tokamak, spherical tokamak (ST), FRC and mirror machine are shown.

\end{abstract}


\end{frontmatter}


\noindent
{\bf PROGRAM SUMMARY}

\begin{small}
\noindent
{\em Program Title:}  BORAY                                      \\
{\em Licensing provisions:}        BSD 3-clause                            \\
{\em Programming language:}    Matlab                               \\
{\em Nature of problem:} Solve the plasmas electromagnetic wave propagation and absorption in the geometrical optics approximation for magnetized plasmas based on ray tracing of plasma dispersion relation. In axisymmetric $(r,z)$ coordinates, the code can be used for both closed and open field lines plasmas of various configurations such as tokamak, spherical tokamak, FRC and mirror machine. \\
{\em Solution method: }  Runge-Kutta time integral to solve ray tracing equations for wave propagation, and integral the imaginary part of the wave frequency in hot kinetic dispersion relation for wave absorption.\\
{\em Additional comments including Restrictions and Unusual features (approx. 50-250 words):} Kinetic relativistic effects are not included in the present version yet. Only axisymmetric two-dimensional (2D) profiles are support in present version.\\

\end{small}


\section{Introduction}

In magnetized confinement plasmas, waves heating is one of the most important approach to heating the plasma to high temperature ($>$10keV). The usually used waves from high frequency ($\sim$100GHz) to low frequency ($<$1MHz) include electron cyclotron wave (ECW), lower hybrid wave (LHW), ion cyclotron wave (ICW)  and Alfv\'en wave (AW). There are also terminologies such as fast wave (FW), slow wave (SW), helicon wave, etc. A simple but still accurate way to study the wave propagation and heating is using the geometrical optics approximation, which yields the ray  tracing equations.

The ray tracing equations in Cartesian coordinates are
\begin{eqnarray}\label{eq:}
\frac{d{\bm r}}{dt}&=&\frac{\partial\omega}{\partial{\bm k}}=-\frac{\partial D/\partial{\bm k}}{\partial D/\partial\omega}={\bm v}_g,\\
\frac{d{\bm k}}{dt}&=&-\frac{\partial\omega}{\partial{\bm r}}=\frac{\partial D/\partial{\bm r}}{\partial D/\partial\omega},
\end{eqnarray}
with the dispersion relation
\begin{eqnarray}\label{eq:DR}
D(\omega,{\bm k},{\bm r})=0,
\end{eqnarray}
where ray position ${\bm r}=(x,y,z)$ and wave vector ${\bm k}=(k_x,k_y,k_z)$. Here, $\omega$ is wave frequency, and ${\bm v}_g$ is wave group velocity.

Several widely used ray tracing codes are available in magnetic confinement fusion community, such as GENRAY\cite{Smirnov2003}, TORAY\cite{Mazzucato1987}, C3PO\cite{Peysson2012}, CURRAY\cite{Brambilla1986}  and TASK/WR\cite{Fukuyama2018}. However, most of them are developed for tokamak and use single fluid magnetohydrodynamics (MHD) equilibrium, thus the density and temperature profiles are set to be magnetic flux functions, and the open field line region is also omitted or simplified. The assumption that the density and temperature profiles be flux functions is helpful to obtain the flux average power absorption and to calculate the current driven. These treatments can be useful and sufficient for study the present day tokamak. However, they can not be used to the configurations with open field line plasmas or when the density and temperature are not magnetic flux functions. There are also codes for some special cases such as RAYS\cite{Batchelor1982} (which is later updated to TORAY) for mirror configuration and FRTC\cite{Esterkin1996} for LHW. In Ref.\cite{Shalashov2012}, a simplified model is used to study the ECRH in mirror. Thus the need for relaxing these restrictions in order to make a code applicable for all situations has motivated our present work.
The present work is an extended version of the fluid and kinetic plasma dispersion relation solver BO code\cite{Xie2019,Xie2016,Xie2014}.

\section{Equations to Solve}\label{sec:boraymodel}

In this work, we use cylindrical coordinates $(r,\phi,z)$. The wave vector variables are chosen as $(k_r,n_\phi=rk_\phi,k_z)$.  The coordinates relations are $x=r\cos\phi$, $y=r\sin\phi$, $k_x=k_r\cos\phi-\frac{n_\phi}{r}\sin\phi$ and $k_y=k_r\sin\phi+\frac{n_\phi}{r}\cos\phi$. Note that the canonical coordinate for $\phi$ is $n_\phi$, not $k_\phi$. If we use $k_\phi$ as a coordinate, the ray tracing equation expressions would be more complicated, c.f., Ref.\cite{McVey1979}.

\subsection{Ray tracing equations in cylindrical coordinates}
Do the coordinates transformation from $(x,y,z,k_x,k_y,k_z)$ to $(r,\phi,z,k_r,n_\phi,k_z)$, we can have
\begin{eqnarray}\label{eq:}
\frac{dr}{d\tau}=\frac{\partial D}{\partial k_r},~
\frac{d\phi}{d\tau}=\frac{\partial D}{\partial n_\phi},~
\frac{dz}{d\tau}=\frac{\partial D}{\partial k_z},~\\
\frac{dk_r}{d\tau}=-\frac{\partial D}{\partial r},~
\frac{dn_\phi}{d\tau}=-\frac{\partial D}{\partial\phi},~
\frac{dk_z}{d\tau}=-\frac{\partial D}{\partial z},~
\end{eqnarray}
with
\begin{eqnarray}\label{eq:}
\frac{dt}{d\tau}=-\frac{\partial D}{\partial \omega},
\end{eqnarray}
Usually, the dispersion relation (\ref{eq:DR}) is written as $D=D(\omega,k_\parallel,k_\perp^2)=0$.
Here, the parallel wave vector $k_\parallel={\bm k}\cdot{\bm b}=\frac{1}{B}\Big(k_rB_r+k_zB_z+\frac{n_\phi}{r}B_\phi\Big)$ is defined from the magnetic field $\bm B$, and $k_\perp^2=k^2-k_\parallel^2$, $B=B(r,z)=\sqrt{B_r^2+B_z^2+B_\phi^2}$, $k^2=k_r^2+k_z^2+\frac{n_\phi^2}{r^2}$.

We only considered axisymmetric configurations, i.e., $\frac{\partial D}{\partial \phi}=0$.
We need to calculate the ray tracing equation from $(r,\phi,z,k_r,n_\phi,k_z)$ to $(r,\phi,z,k_\parallel^2,k_\perp^2,\alpha)$ with $\frac{\partial D}{\partial \alpha}=0$, where $\alpha$ is the angle relevant to two perpendicular wave vectors and can be omitted here. We obtain
\begin{eqnarray}\label{eq:}
\frac{\partial D}{\partial k_r}\Big|_{r,\phi,z,n_\phi,k_z}=2\Big(\frac{\partial D}{\partial k_\parallel^2}-\frac{\partial D}{\partial k_\perp^2}\Big)k_\parallel\frac{B_r}{B}+2\frac{\partial D}{\partial k_\perp^2}k_r,\\
\frac{\partial D}{\partial n_\phi}\Big|_{r,\phi,z,k_r,k_z}=2\Big(\frac{\partial D}{\partial k_\parallel^2}-\frac{\partial D}{\partial k_\perp^2}\Big)k_\parallel\frac{B_\phi}{rB}+2\frac{\partial D}{\partial k_\perp^2}\frac{n_\phi}{r^2},\\
\frac{\partial D}{\partial k_z}\Big|_{r,\phi,z,k_r,n_\phi}=2\Big(\frac{\partial D}{\partial k_\parallel^2}-\frac{\partial D}{\partial k_\perp^2}\Big)k_\parallel\frac{B_z}{B}+2\frac{\partial D}{\partial k_\perp^2}k_z,\\\nonumber
\frac{\partial D}{\partial r}\Big|_{\phi,z,k_\parallel^2,k_\perp^2,\alpha}=\frac{\partial D}{\partial r}\Big|_{\phi,z,k_r,n_\phi,k_z}\\
~~~+2\Big(\frac{\partial D}{\partial k_\parallel^2}-\frac{\partial D}{\partial k_\perp^2}\Big)k_\parallel\frac{\partial k_\parallel}{\partial r}-2\frac{\partial D}{\partial k_\perp^2}\frac{n_\phi^3}{r^3}\\
\frac{\partial D}{\partial \phi}\Big|_{r,z,k_r,n_\phi,k_z}=0,\\
\frac{\partial D}{\partial z}\Big|_{r,\phi,k_r,n_\phi,k_z}=\frac{\partial D}{\partial r}\Big|_{\phi,z,k_\parallel^2,k_\perp^2,\alpha}+2\Big(\frac{\partial D}{\partial k_\parallel^2}-\frac{\partial D}{\partial k_\perp^2}\Big)k_\parallel\frac{\partial k_\parallel}{\partial z},
\end{eqnarray}
where
\begin{eqnarray}\label{eq:}
\frac{\partial k_\parallel}{\partial r}=-\frac{k_\parallel}{B}\frac{\partial B}{\partial r}+\frac{1}{B}\Big(k_r\frac{\partial B_r}{\partial r}+k_z\frac{\partial B_z}{\partial r}+\frac{n_\phi}{r}\frac{\partial B_\phi}{\partial r}-\frac{B_\phi n_\phi}{r^2}\Big),\\
\frac{\partial k_\parallel}{\partial z}=-\frac{k_\parallel}{B}\frac{\partial B}{\partial z}+\frac{1}{B}\Big(k_r\frac{\partial B_r}{\partial z}+k_z\frac{\partial B_z}{\partial z}+\frac{n_\phi}{r}\frac{\partial B_\phi}{\partial z}\Big).
\end{eqnarray}

\subsection{Ray tracing equations for cold plasma dispersion relation}
The cold plasma dispersion relation is
\begin{eqnarray}\label{eq:colddr}\nonumber
F(\omega,k_\parallel^2,k_\perp^2)=\varepsilon_1\frac{k_\perp^4c^4}{\omega^4}-\Big[(\varepsilon_1+\varepsilon_3)\Big(\varepsilon_1-\frac{k_\parallel^2c^2}{\omega^2}\Big)-\varepsilon_2^2\Big]\\
\frac{k_\perp^2c^2}{\omega^2}+\varepsilon_3\Big[\Big(\varepsilon_1-\frac{k_\parallel^2c^2}{\omega^2}\Big)^2-\varepsilon_2^2\Big]=0,
\end{eqnarray}
where
\begin{eqnarray}\label{eq:}\nonumber
\varepsilon_1=1-\sum_s\frac{\omega_{ps}^2}{\omega^2-\omega_{cs}^2},~~\varepsilon_2=\sum_s\frac{\omega_{cs}}{\omega}\frac{\omega_{ps}^2}{\omega^2-\omega_{cs}^2},~~\\
\varepsilon_3=1-\sum_s\frac{\omega_{ps}^2}{\omega^2},\\
{\bm n}=\frac{{\bm k}c}{\omega},~\omega_{cs}=\frac{q_sB}{m_s},~\omega_{ps}=\sqrt{\frac{n_{s0}q_s^2}{\epsilon_0m_s}},~~c=\frac{1}{\sqrt{\mu_0\epsilon_0}}.
\end{eqnarray}
The derivates expressions for  $F$ can be readily written out explicit, i.e., $\frac{\partial F}{\partial k_\parallel^2}$, $\frac{\partial F}{\partial k_\perp^2}$, $\frac{\partial F}{\partial r}\Big|_{\phi,z,k_\parallel^2,k_\perp^2,\alpha}$, $\frac{\partial F}{\partial z}\Big|_{r,\phi,k_\parallel^2,k_\perp^2,\alpha}$ and $\frac{\partial F}{\partial \omega}$, which are not shown here.

We need the 2D equilibrium profiles $B_r(r,z)$, $B_z(r,z)$, $B_\phi(r,z)$, $B_(r,z)$, $n_{s0}(r,z)$ and their first order derivate $\partial/\partial r$ and $\partial/\partial z$. 

We should note that the cold plasma dispersion relation Eq.(\ref{eq:colddr}) is singular at cyclotron resonant position of species $s=a$, say, if $\omega\simeq\omega_{c,s=a}$.  This singularity is first order to $Y_a=1-\omega_{c,s=a}^2/\omega^2$, thus we can multiple Eq.(\ref{eq:colddr}) to define a new dispersion relation $G(\omega,k_\parallel^2,k_\perp^2)=Y_aF(\omega,k_\parallel^2,k_\perp^2)$ in the code, which is similar to the treatment in GENRAY\cite{Smirnov2003}.  One should also be careful that we do not use $G'(\omega,k_\parallel^2,k_\perp^2)=Y_a^nF(\omega,k_\parallel^2,k_\perp^2)$ ($n\geq2$), which will cause the group velocity vanish at resonant position.

It should be noted that the above mathematical formulation is pertinent to both open and closed field line plasmas, and therefore the equations are applicable for both.

\subsection{Wave absorption equation using hot kinetic plasma dispersion relation}
Since the drift bi-Maxwellian distribution function may lead unstable modes with imaginary part of wave frequency be positive, i.e., wave absorbs energy from particles, we only use the Maxwellian distribution hot kinetic dispersion relation for wave heating in the present version of BORAY. The non-relativistic dispersion tensor is standard, c.f., \cite{Xie2016}.

For weak damping approximation with $\omega_i\ll\omega_r$ and ${\bm k}_i\ll{\bm k}_r$, for $D(\omega,{\bm k})=D_r(\omega,{\bm k})+iD_i(\omega,{\bm k})=0$, $D_i\ll D_r$, we have $D_r(\omega_r,{\bm k}_r)=0$ and
\begin{eqnarray}\label{eq:}
i\Big[\frac{\partial D_r(\omega_r,{\bm k}_r)}{\partial\omega_r}\omega_i+D_i(\omega_r,{\bm k}_r)\Big]\simeq0,\\
i\Big[\frac{\partial D_r(\omega_r,{\bm k}_r)}{\partial  {\bm k}_r}{\bm k}_i+D_i(\omega_r,{\bm k}_r)\Big]\simeq0,
\end{eqnarray}
we have
\begin{eqnarray}\label{eq:kiwi}
{\bm k}_i=-\frac{D_i}{\partial D_r/\partial {\bm k}_r},~~\omega_i=-\frac{D_i}{\partial D_r/\partial \omega_r},\\
{\bm k}_i=\frac{\partial D_r/\partial\omega_r}{\partial D_r/\partial {\bm k}_r}\omega_i=-\frac{\omega_i}{\partial\omega/\partial {\bm k}_r}=-\frac{\omega_i}{{\bm v}_g},
\end{eqnarray}
So the wave damping caused wave intensity $P(s)$ along the trajectory can be calculated from either ${\bm k}_i$ or $\omega_i$, due to
\begin{eqnarray}\label{eq:}\nonumber
P(s)=P_0\cdot e^{-2\int_0^s{\bm k}_{i}\cdot d{\bm r}}=P_0\cdot e^{-2\int_0^t\omega_i dt}.
\end{eqnarray}

One can use Eq.(\ref{eq:kiwi}) to calculate $\omega_i$ when the weak damping assumption is valid. In BORAY, we choose to calculate the complex $\omega=\omega_r+i\omega_i$ more accurately along the ray, since that usually the kinetic $\omega_r$ may deviate from initial wave frequency $\omega_0$.
The BO code\cite{Xie2019,Xie2016,Xie2014} is convenient to calculate the $\omega$ for given real $\bm k$ for either kinetic or fluid plasma dispersion relations. So, we succeed the corresponding modules of BO to calculate the kinetic absorption in the present ray tracing code. Hence, we named the present code as BO-RAY (or, BORAY) as a branch of BO family.

To calculate the absorption ratio from different species, we keep only the temperature of that species un-change and set the temperatures of other species to be cold.

\section{Benchmarks and Applications}
In this section, to show the accuracy and capability of BORAY, we compare it with other ray tracing code, particle simulation and experiments. If not specialized, the tokamak and ST equilibria in the following examples are obtained from EFIT\cite{Lao1985} outputs of corresponding configurations. All the examples in this section are summarized in Table \ref{tab:examplessummary}, which are obtained by BORAY under a unified theoretical model and numerical code as described in Sec.\ref{sec:boraymodel}, i.e., we do not need to choose different models for different examples. The only differences between different examples are the input magnetic fields, densities and temperatures profiles, and the initial wave frequency, position and wave vector.

\begin{table}[htp]
\caption{BORAY benchmark and application examples, for varies wave frequencies and varies configurations, with numerical MHD equilibria (default), 3-fluid equilibrium (Fig.\ref{fig:EXL_28GHz_ECW} ) and analytical MHD equilibrium (Fig.\ref{fig:EXL_4_5MHz_HHe_ICW}).}
\begin{center}
{\scriptsize
\begin{tabular}{c|cccc}\hline
Configuration& Tokamak  & ST & FRC & Mirror \\
(Field Lines) & (Closed)  & (Both) &  (Both) & (Open)\\
$B_\phi$ & $\neq0$  & $\neq0$ &  $=0$ & $=0$ \\\hline
ECW (O\&X) & Fig.\ref{fig:EAST_100GHz_ECW} & Fig.\ref{fig:EXL_28GHz_ECW} (3-fluid eq.) &  & 
\\\hline

LHW & Fig.\ref{fig:EAST_2_45GHz_LHW} &  &  & Fig.\ref{fig:mirror_160MHz_LHW}
\\\hline

Helicon & Fig.\ref{fig:HL2M_476MHz_Helicon} &  &  & 
\\\hline

HHFW &  & & Fig.\ref{fig:C2U_7MHz_HHFW}  &
\\\hline

ICW & & Fig.\ref{fig:EXL_5MHz_ICW},  &  & Fig.\ref{fig:KMAX_750kHz_ICW}\\

 & & Fig.\ref{fig:EXL_4_5MHz_HHe_ICW} (analy. eq.)  &  &
\\\hline

\end{tabular}}
\end{center}
\label{tab:examplessummary}

\end{table}%

\begin{figure}
\centering
\includegraphics[width=8cm]{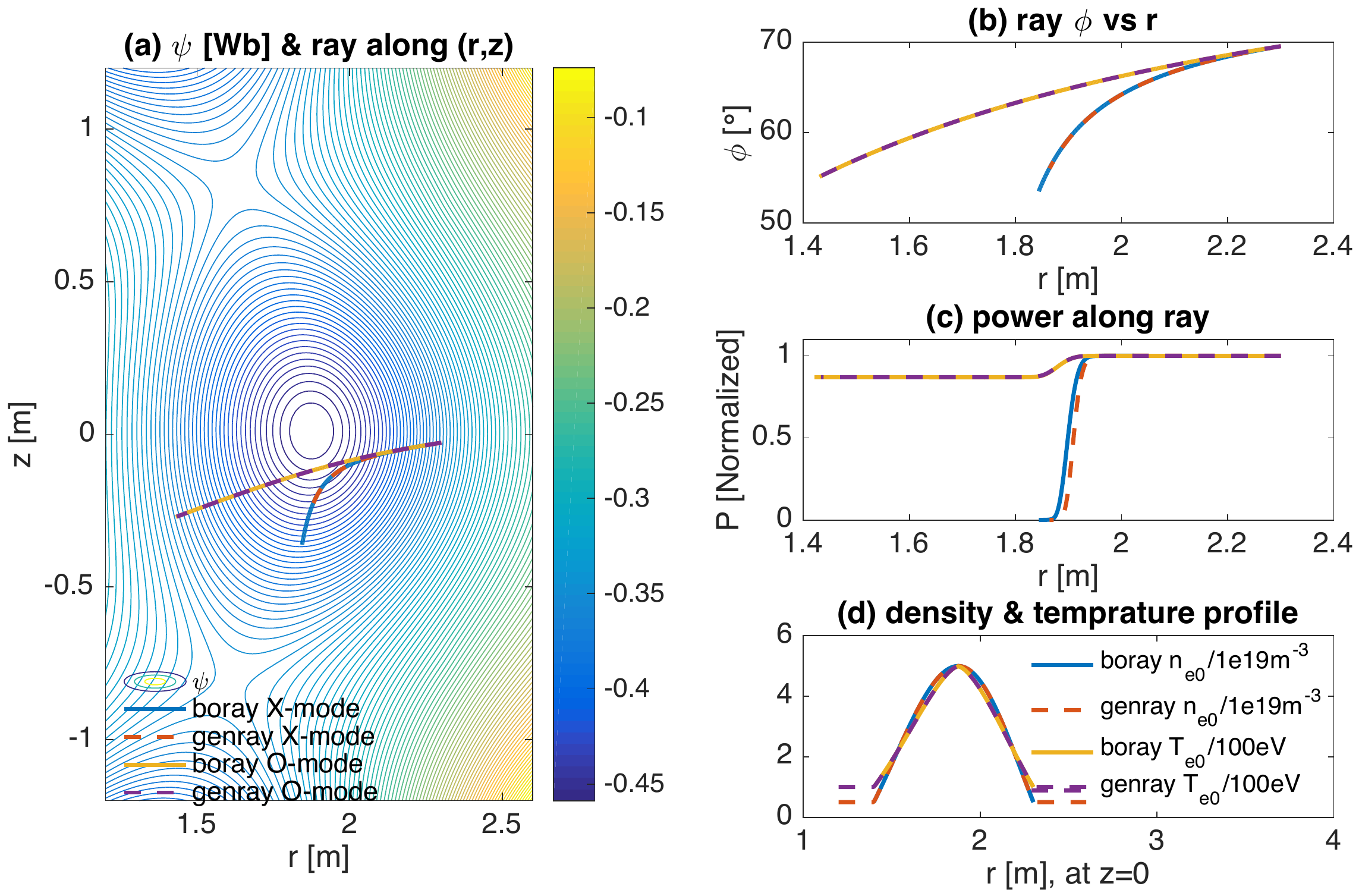}\\
\caption{Comparison of BORAY and  GENRAY for EAST tokamak 100GHz ECW O and X modes. Both ray trajectories and power absorptions agree well.}\label{fig:EAST_100GHz_ECW}
\end{figure}

\begin{figure}
\centering
\includegraphics[width=8cm]{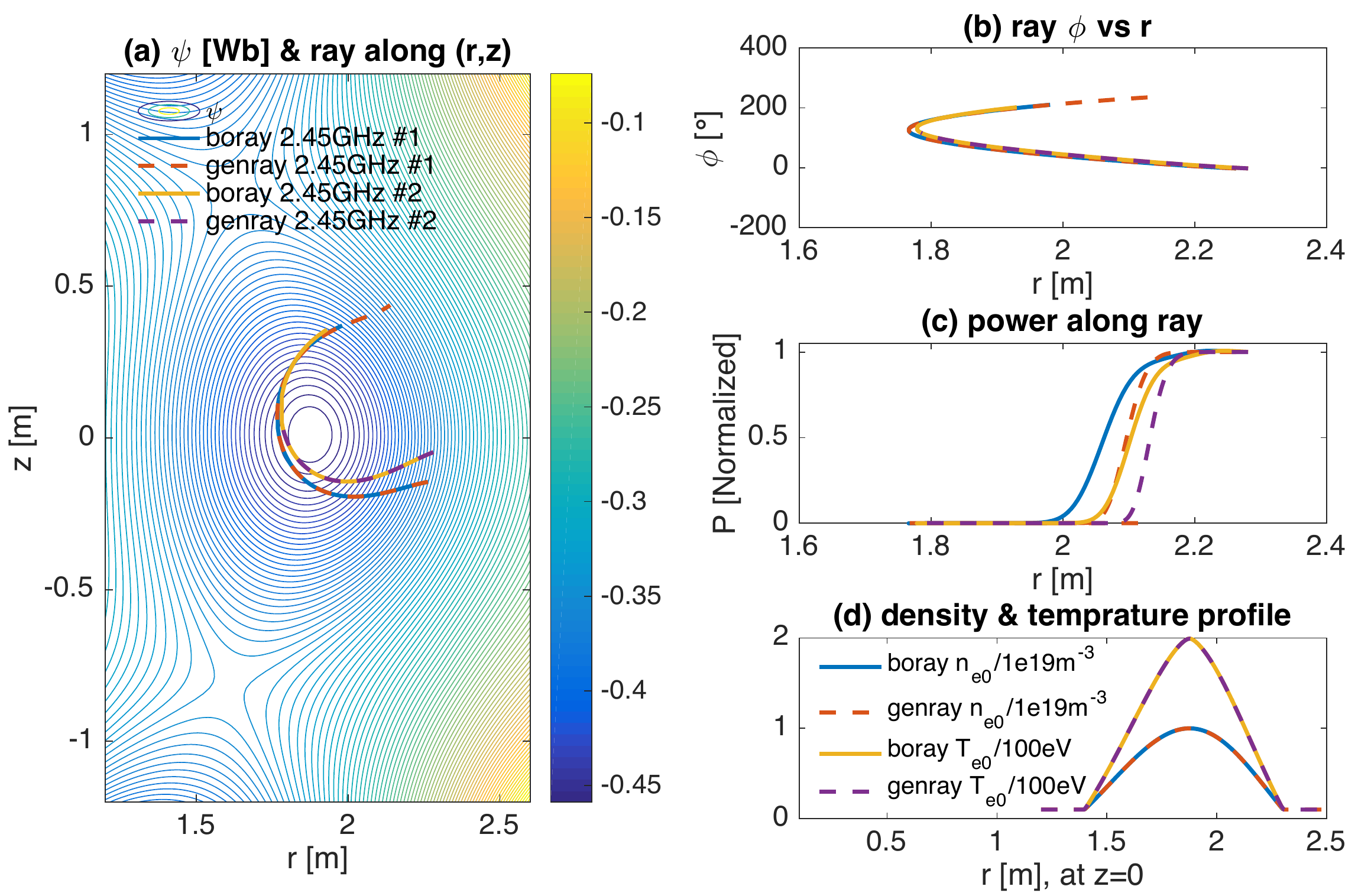}\\
\caption{Comparison of BORAY and  GENRAY for EAST tokamak 2.45GHz LHW. Ray trajectories agree well.  However, GENRAY damping early than BORAY for power absorptions, which may due to different absorption models used in the two codes.}\label{fig:EAST_2_45GHz_LHW}
\end{figure}

\begin{figure}
\centering
\includegraphics[width=8cm]{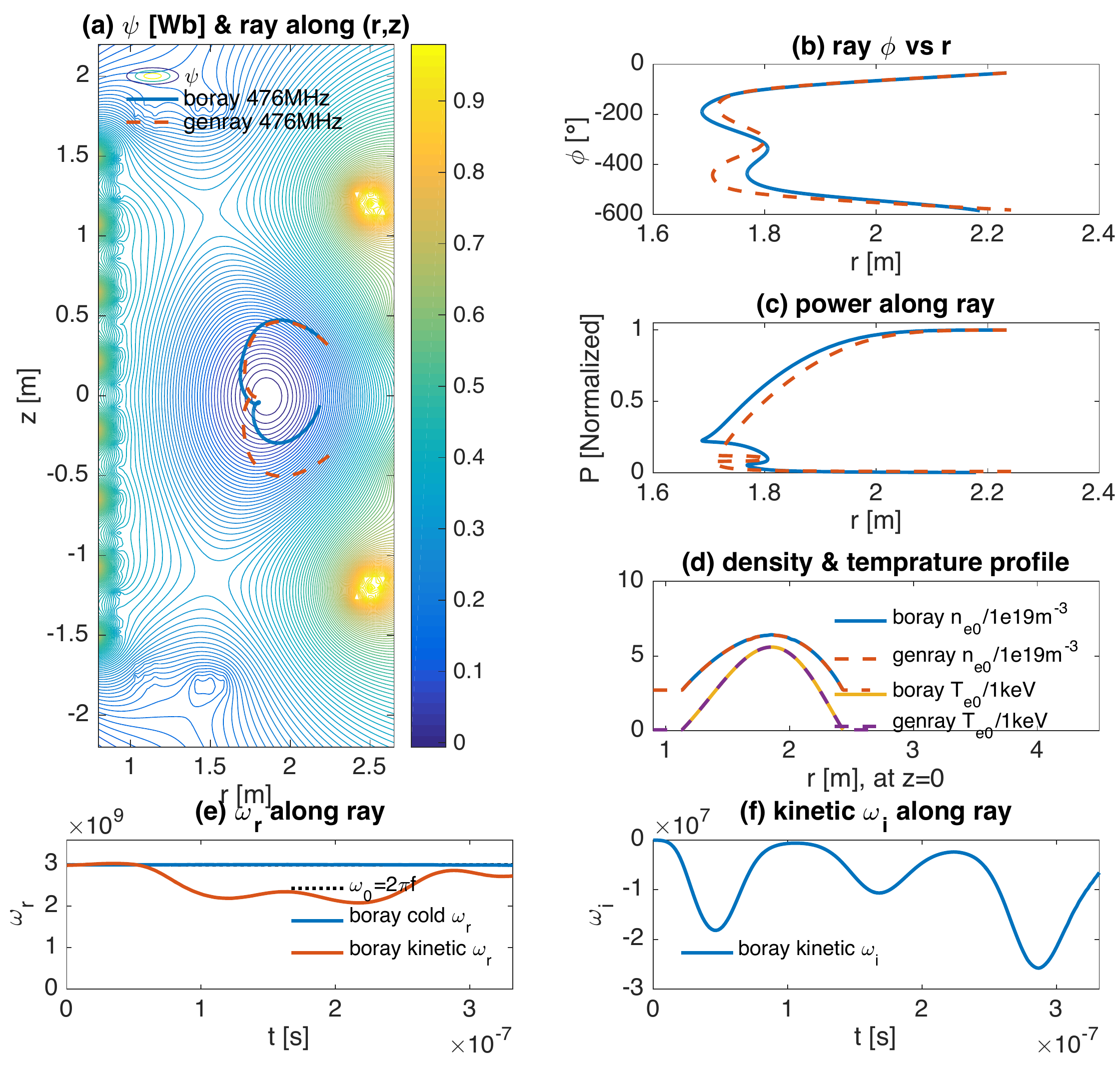}\\
\caption{Comparison of BORAY and  GENRAY for HL-2M tokamak 476MHz helicon wave\cite{Li2020}. Ray trajectories and power absorptions roughly agree. The difference may come from the numerical errors in GENRAY, since less than 200 points is used along the ray in GENRAY output.}\label{fig:HL2M_476MHz_Helicon}
\end{figure}

\begin{figure}
\centering
\includegraphics[width=8cm]{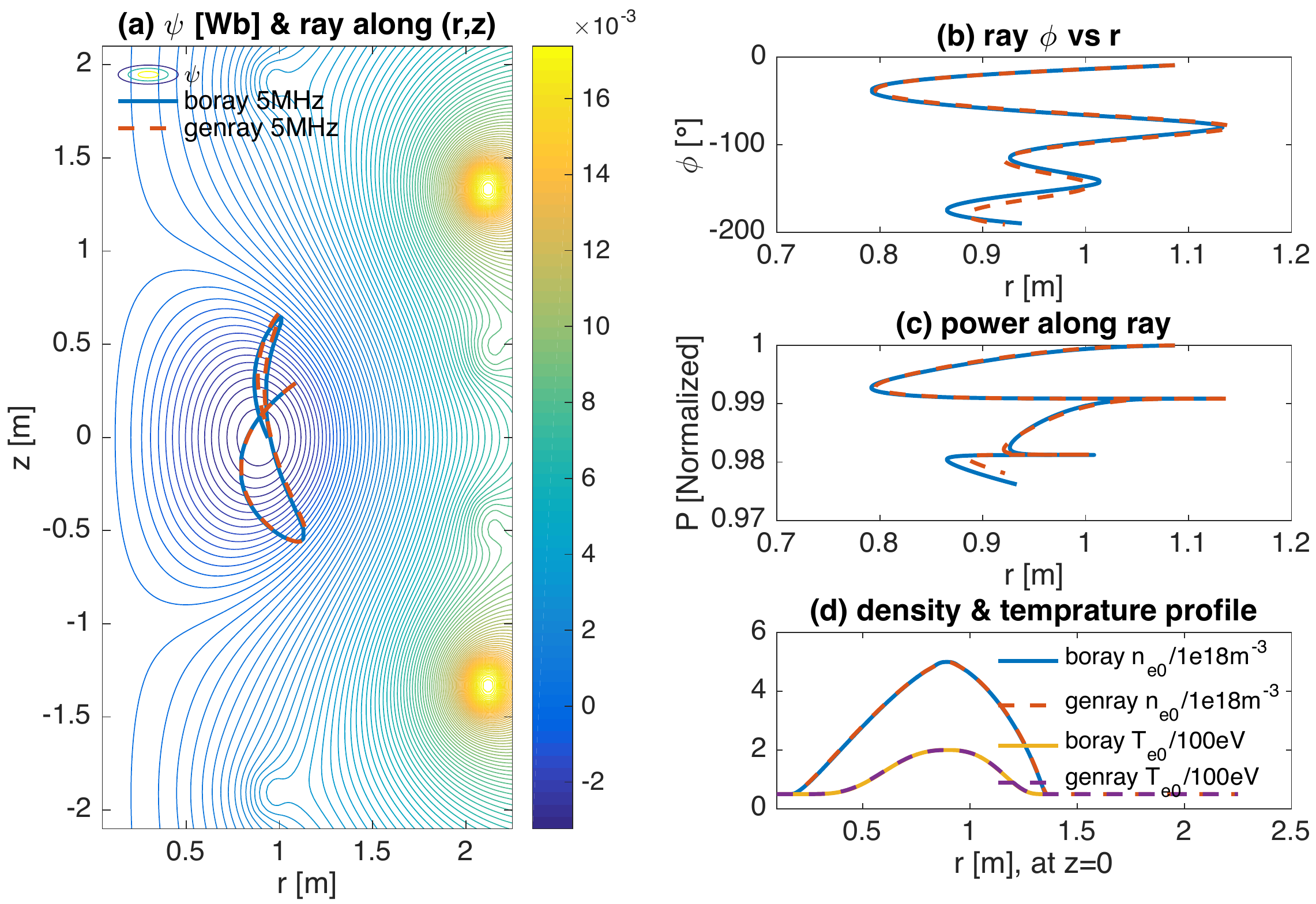}\\
\caption{Comparison of BORAY and GENRAY for EXL-50 spherical tokamak 5MHz ICW. Both ray trajectories and power absorptions agree well. The slight difference may come from numerical error of time push or grid interpolation.}\label{fig:EXL_5MHz_ICW}
\end{figure}

\subsection{Tokamak and ST ECW, LHW and ICW}

In this subsection, we show the benchmarks between BORAY and GENRAY for several standard tokamak and ST cases, including the frequency from high to low, i.e., ECW X-mode and O-mode, LHW, helicon wave, and ICW.

Fig.\ref{fig:EAST_100GHz_ECW} shows the comparison of BORAY and  GENRAY for EAST tokamak ECW X-mode and O-mode, with central magnetic field $B_0=1.78T$,  major radius $R_0=1.88m$, safety factor $q_0=1.5$, density $n_{e0}=5\times10^{19}m^{-3}$ and temperature $T_{e0}=T_{i0}=500eV$. 

Fig.\ref{fig:EAST_2_45GHz_LHW} shows the comparison of BORAY and GENRAY for EAST tokamak 2.45GHz LHW for the same equilibrium magnetic fields as in Fig.\ref{fig:EAST_100GHz_ECW}, but different densities and temperatures, $n_{e0}=1\times10^{19}m^{-3}$ and $T_{e0}=T_{i0}=200eV$.

Fig.\ref{fig:HL2M_476MHz_Helicon} shows the comparison of BORAY and  GENRAY for HL-2M tokamak 476MHz helicon wave, with central magnetic field $B_0=1.76T$, major radius $R_0=1.85m$, safety factor $q_0=0.98$, density $n_{e0}=6.42\times10^{19}m^{-3}$, $n_{D^{+}0}=5.39\times10^{19}m^{-3}$, $n_{C^{+6}0}=0.17\times10^{19}m^{-3}$ and temperature $T_{e0}=5.60keV$, $T_{D^{+}0}=T_{C^{+6}0}=4.78keV$.

Fig.\ref{fig:EXL_5MHz_ICW} shows the comparison of BORAY and  GENRAY for EXL-50 spherical tokamak 5MHz ICW, with central magnetic field $B_0=0.26T$, major radius $R_0=0.89m$, safety factor $q_0=10.9$, density $n_{e0}=5.0\times10^{18}m^{-3}$ and temperature $T_{e0}=200eV$ and $T_{i0}=50eV$. 

In all these above benchmark cases, both ray trajectories and power absorptions agree well. Some differences may come from numerical error or slight different models implemented in the two codes. We carry out analysis to check validity of BORAY results. For example, in Fig.\ref{fig:HL2M_476MHz_Helicon}(e), we show the $\omega_r$ along the ray, which are solved from the cold and kinetic dispersion relations with the ray tracing output $\bm k$ along the ray. We see that the cold plasma $\omega_r$ is almost identical to the given input wave frequency $\omega_0=2\pi{}f$, which means that the cold plasma ray tracing equation is solved accurately in BORAY. The deviation of kinetic $\omega_r$ to $\omega_0$ implies that the cold plasma assumption for the ray tracing may not be accurate for this case. However, the $\omega_i/\omega_r\simeq0.01$ means that the weak damping assumption still holds.

\subsection{FRC High Harmonic Fast Wave}
\begin{figure}
\centering
\includegraphics[width=8cm]{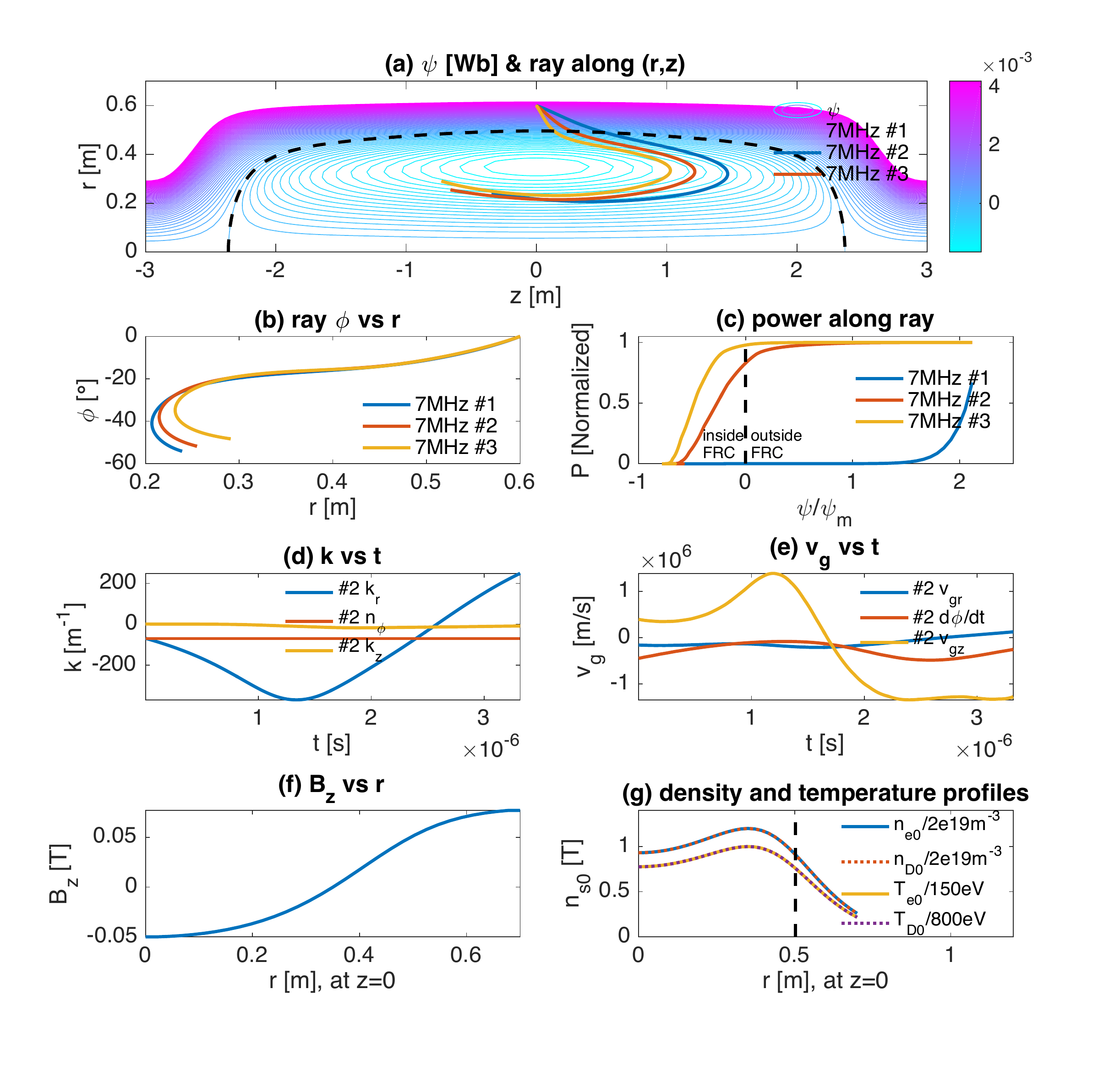}\\
\caption{C2-U FRC 7MHz HHFW simulation results, which is similar to the results in Ref.\cite{Yang2017}, i.e., the  absorption can be 100\% and most power can be deposited inside the closed flux surface for optimized wave parameters.}\label{fig:C2U_7MHz_HHFW}
\end{figure}

A major advantage of BORAY over other ray tracing codes is that BORAY can support both closed and open field lines plasmas equally as defaults. We firstly show the result for FRC case.
Fig.\ref{fig:C2U_7MHz_HHFW}  shows C2-U FRC 7MHz High Harmonic FW (HHFW) simulation results, which is similar to the GENRAY-C results in Ref.\cite{Yang2017}, i.e., the  absorption can be 100\% and most power can be deposited inside the closed flux surface for optimized wave parameters. The equilibrium is generated by GSEQ-FRC\cite{Ma2021} using similar parameters as in Ref.\cite{Yang2017}, with axis magnetic field $B(0,0)=-0.05T$,  major radius $R_0=0.35m$,  central density $n_{e0}=2.4\times10^{19}m^{-3}$ and temperature $T_{e0}=150eV$ and $T_{D^{+1}0}=800eV$.

\begin{figure}
\centering
\includegraphics[width=8cm]{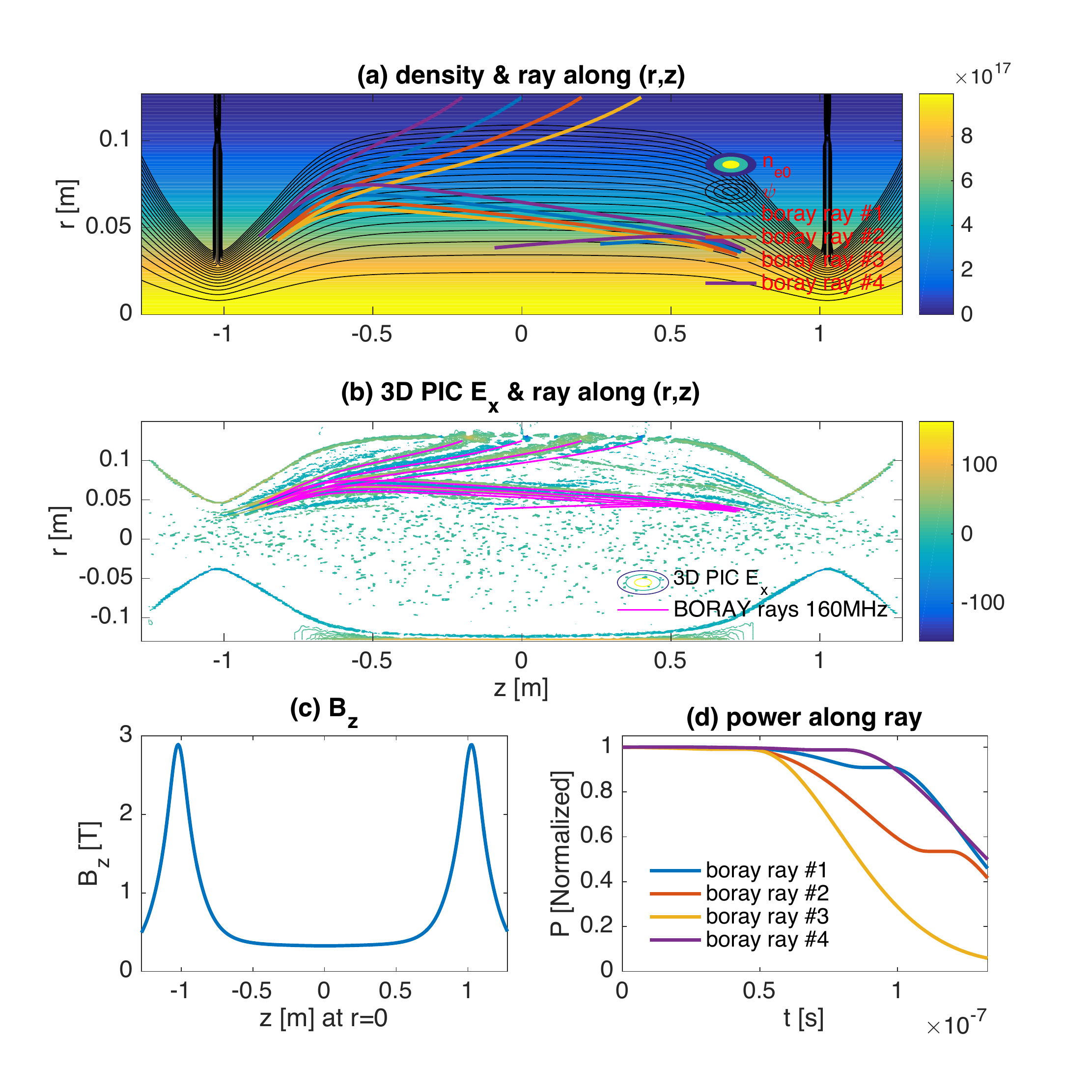}\\
\caption{Mirror 160MHz LHW simulation results, which is close to the results in Ref.\cite{Zheng2021}  of 3D PIC simulations (b), where the PIC data with $E_x<10$ is removed to make the figure more clear. The power absorptions (d)  are also similar, i.e., the wave is almost decayed away before reaching the second turning point.}\label{fig:mirror_160MHz_LHW}
\end{figure}

\begin{figure}
\centering
\includegraphics[width=8cm]{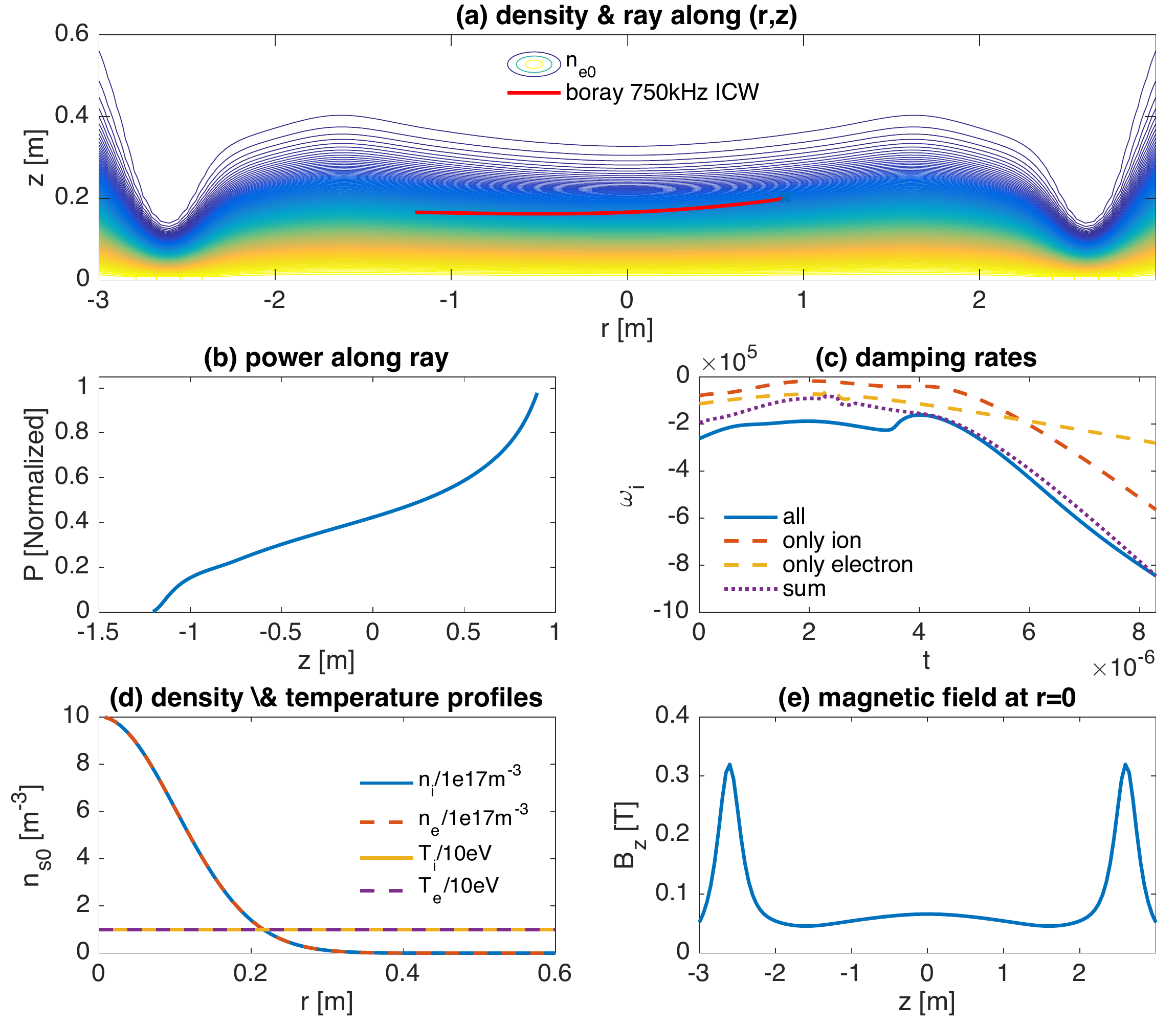}\\
\caption{KMAX mirror 750kHz ICW simulation results, which is close to the results in Ref.\cite{Liu2018}  of ICRF experiment, i.e., absorption rate $>40\%$.}\label{fig:KMAX_750kHz_ICW}
\end{figure}

\subsection{Mirror LHW and ICW}
Here, we show the capability of BORAY for mirror configuration.   
Fig.\ref{fig:mirror_160MHz_LHW} shows mirror 160MHz LHW simulation results, which is close to the three dimensional (3D) electromagnetic particle-in-cell (PIC) simulations results in Ref.\cite{Zheng2021} as shown in Fig.\ref{fig:mirror_160MHz_LHW}(b) for both the ray trajectories and turning points. Here, the initial antenna $k_z=16.3m^{-1}$. Note that the density is given as $n_{s0}=n_0e^{-\frac{r^2}{2\sigma^2}}$, which is not set as function of magnetic flux. Here, $n_0=1\times10^{18}m^{-3}$ and $\sigma=0.045m$. Temperature is set as constant for both electrons and H ions, $T_{e0}=T_{H0}=460eV$.

Fig.\ref{fig:KMAX_750kHz_ICW} shows KMAX mirror 750kHz ICW simulation results, which is close to the results in Ref.\cite{Liu2018}  of ICRF experiment, i.e., absorption rate $>40\%$. Initial wave parameters $(r,\phi,z,k_{r,guess},n_\phi,k_z)=(0.2,0,0.9,90.2,-1,-8)$. Note that the summation of ion and electron damping rates is not equal to the total damping rate, which is probably due to the violation of weak damping approximation or that the effects of different species are not independently.

\begin{figure}
\centering
\includegraphics[width=8cm]{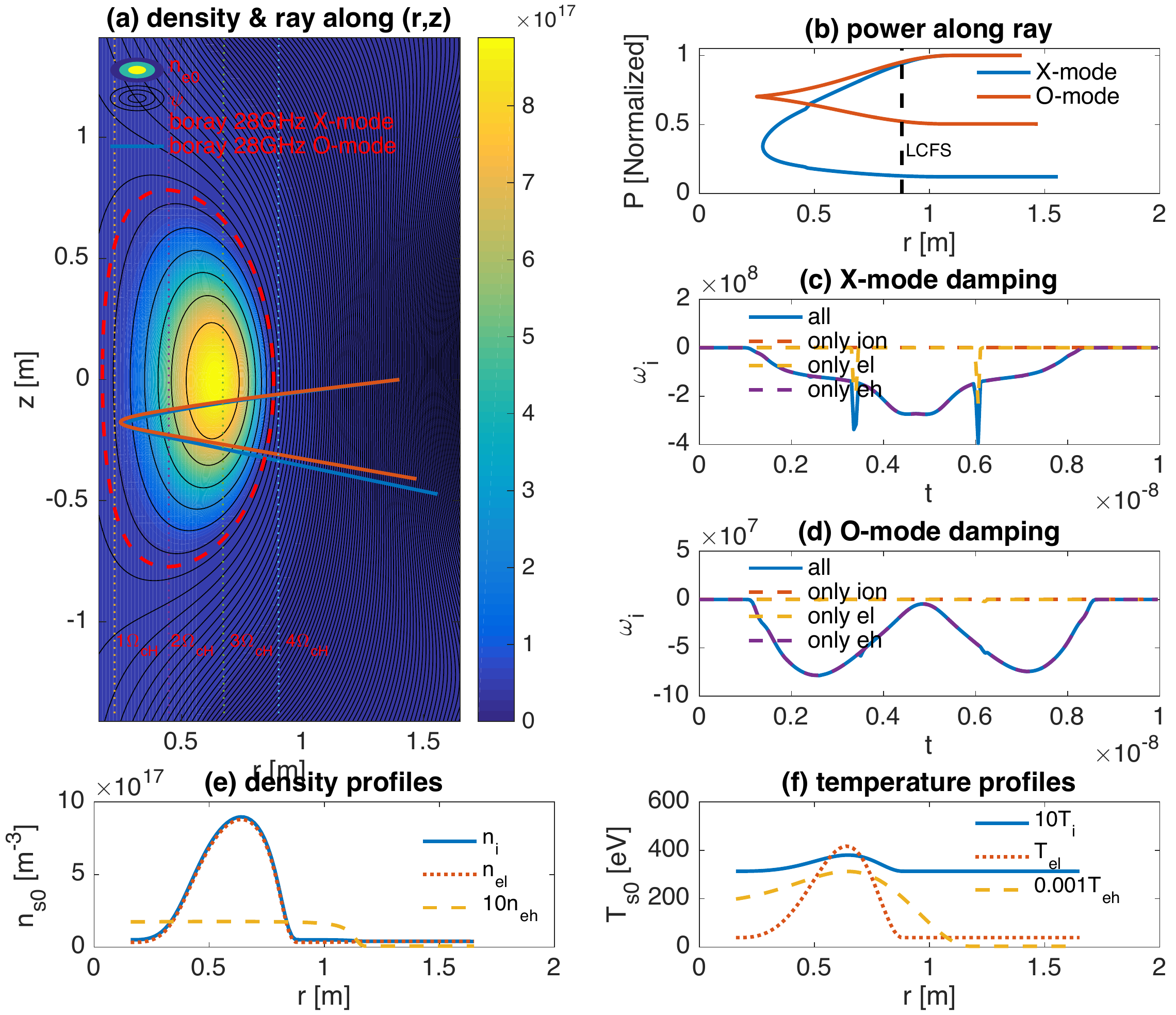}\\
\caption{EXL-50 spherical tokamak 28GHz ECW O\&X-modes, with three-fluid equilibrium. The high energy electrons ($eh$) contribute most of the power absorption, whereas the absorption from ions and low energy electrons ($el$) are negligible.}\label{fig:EXL_28GHz_ECW}
\end{figure}

\subsection{ST ECW under multi-fluid equilibrium}
In some STs, such as QUEST\cite{Onchi2021} and EXL-50\cite{Shi2021}, the high energy electrons ($>$10keV) are an important component. Ref.\cite{Ishida2021} provides a multi-fluid equilibrium model for EXL-50. Here, we show the capability of BORAY for this equilibrium configuration.   
Fig.\ref{fig:EXL_28GHz_ECW} shows the EXL-50 spherical tokamak 28GHz ECW O\&X-modes under three-fluid equilibrium with central magnetic field $B_0=0.36T$, major radius $R_0=0.63m$, safety factor $q_0\simeq10$, maximum densities $n_{H^+}=9.0\times10^{17}m^{-3}$, $n_{el}=8.8\times10^{17}m^{-3}$, $n_{eh}=1.76\times10^{16}m^{-3}$ and temperatures $T_{H^+}=38eV$, $T_{el}=417eV$ and $T_{eh}=313keV$. The low density ($\sim$2\%) high energy electrons ($eh$) contribute most of the power absorption, whereas the absorption from H$^+$ ions and low energy thermal electrons ($el$) are negligible. We obtain the three fluid equilibrium profiles for both inside and outside the last closed flux surface (LCFS) from the model in Ref.\cite{Ishida2021} for EXL-50 shot\#6935 t=4.45s, with total plasma current $I_p=120kA$. We also see that the X-mode have better absorption than O-mode, and the second order $2\omega_{cH}$ resonant is also stronger than that of O-mode. These high energy electron effects are similar to the  recently reported\cite{Ono2020} QUEST experimental and theoretical analysis results. We can also see from Fig.\ref{fig:EXL_28GHz_ECW} (b) that some amount ($\sim5\%$) of the wave absorption is outside the LCFS.

\begin{figure}
\centering
\includegraphics[width=8cm]{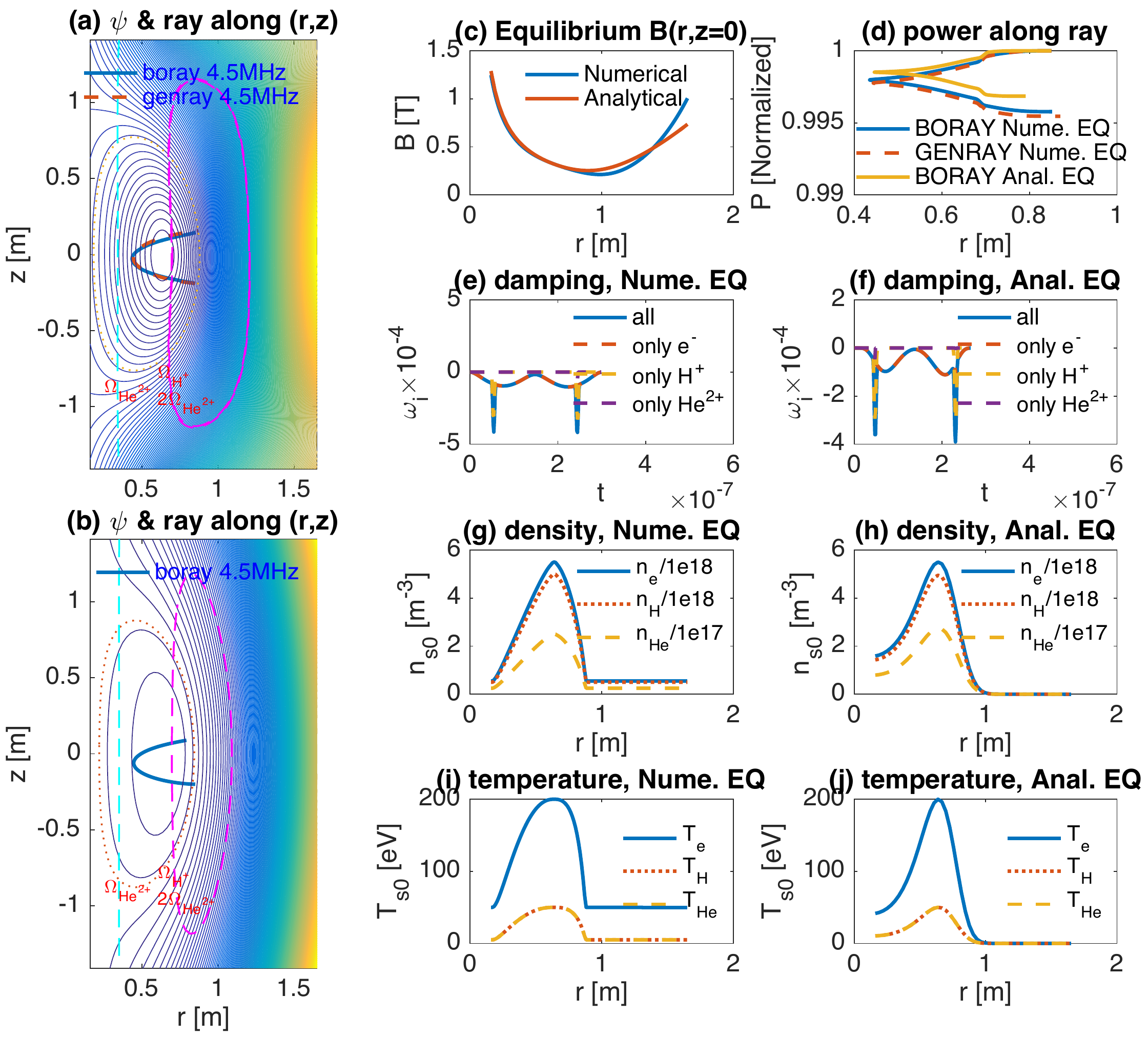}\\
\caption{Comparison EXL-50 spherical tokamak 4.5MHz ICW for numerical and analytical MHD equilibria, with three species, i.e., electrons, H$^{+}$ ions and 5\% He$^{2+}$ minority ions. Both ray trajectories and power absorptions are similar.}\label{fig:EXL_4_5MHz_HHe_ICW}
\end{figure}

\subsection{Comparison of ST ICW under numerical and analytical equilibria}
We are interested to quantify how the characteristics of wave propagation and value of absorption differ between cases of numerically reconstructed and analytical equilibria.
Fig.\ref{fig:EXL_4_5MHz_HHe_ICW} shows the comparison of EXL-50 spherical tokamak 4.5MHz ICW for numerical and analytical MHD equilibria, with three species, i.e., electrons, H$^{+}$ ions and 5\% He$^{2+}$ minority ions. The numerical equilibrium parameters are $B_0=0.32T$, major radius $R_0=0.64m$, safety factor $q_0=1.6$, density $n_{e0}=5.5\times10^{18}m^{-3}$ and temperature $T_{e0}=200eV$ and $T_{i0}=50eV$. The construction of analytical equilibrium is described at \ref{sec:solovev}, with other model parameters $R_x=0.17m$, $E=1.5$, $\tau=0.8$, $L_{ns}=0.9$ and $L_{ts}=0.8$ for the present case. Both ray trajectories and power absorptions are similar for the numerical and analytical equilibria. For both cases, the CPU runtimes of ray tracing are in seconds for 10000 points. The numerical equilibrium case (3s) is slightly faster than the analytical equilibrium case (7s). For this case, most power is absorbed by electrons and fundamental $\Omega_{cH^{+}}$ resonant, with also slight $2\Omega_{cHe^{2+}}$ resonant as can be seen from the damping rate sub-figures (e)\&(f). This benchmark implies that the analytical equilibrium can be good enough to study the rough wave features of the actual devices. 

\section{Summary}
A new plasma wave ray tracing code BORAY (https://github.com/hsxie/boray) has been developed for axisymmetric configurations to support both closed and open field lines plasmas configurations. The code shows good agreement with GENRAY code for tokamak and ST cases of ECW, LHW, helicon wave and ICW, and also agrees well with 3D PIC simulation of LHW in mirror machine, and agrees with GENRAY-C for HHFW in FRC, and  ICW for KMAX mirror experiment. Thus, it can be expected that BORAY can have widely application for the plasma wave propagation and heating studies and especially to help the  design of the wave heating system to choose the wave parameters. The code works for both numerical and analytical equilibria. Future works can include relativistic and collisional effects and calculating the current driven. Modified the cold plasma ray tracing model to kinetic dispersion relation to support electron and ion Bernstein waves could also be an important future topic.

{\it Acknowledgments}
Discussion with Shao-dong Song, Guang-hui Zhu are acknowledged. We also thank Jiang-shan Zheng for providing the 3D PIC mirror LHW simulation data, and Wen-jun Liu for providing the three fluid equilibrium of EXL-50 spherical tokamak. We are grateful to Yu.V. Petrov and R.W. Harvey of CompX for introducing us the details of GENRAY code. This work is supported by the compact fusion project in ENN group.

\appendix

\section{More Details of BORAY}\label{sec:boray}
Bi-linear interpolation is used for uniform $(r,z)$ grids, which can be fast, and even can be faster than analytical equilibrium if we calculated the interpolation coefficients beforehand. In many tests, we find it is accurate enough. For wave absorption, we do not need calculate every point along the ray. Instead, we calculated the ray trajectory firstly with  high accuracy, say $\geq10000$ points, and  then select several points, say 200-1000 points, to calculate the damping rates, and then integral them to obtain the  power absorption. SI units are used for all variables, except that the temperature unit is eV.

The user should generate the initial 2D $(r,z)$ magnetic fields, densities and temperatures profile firstly and also give their derivatives to $r$ and $z$. Also, the user should give the initial wave parameters, i.e., wave frequency $f=\omega/2\pi$ and $(r,\phi,z,k_{r,guess},n_\phi,k_z)$. To make $D(\omega,{\bm k})=0$, BORAY calculate $k_r$ from given $n_\phi$ and $k_z$. Multi-$k_r$ may exist, the user can adjust $k_{r,guess}$ to solve the corresponding $k_r$ who wants. For examples, we use different $k_{r,guess}$ to obtain $X$ and $O$ modes.

\begin{figure}
\centering
\includegraphics[width=8cm]{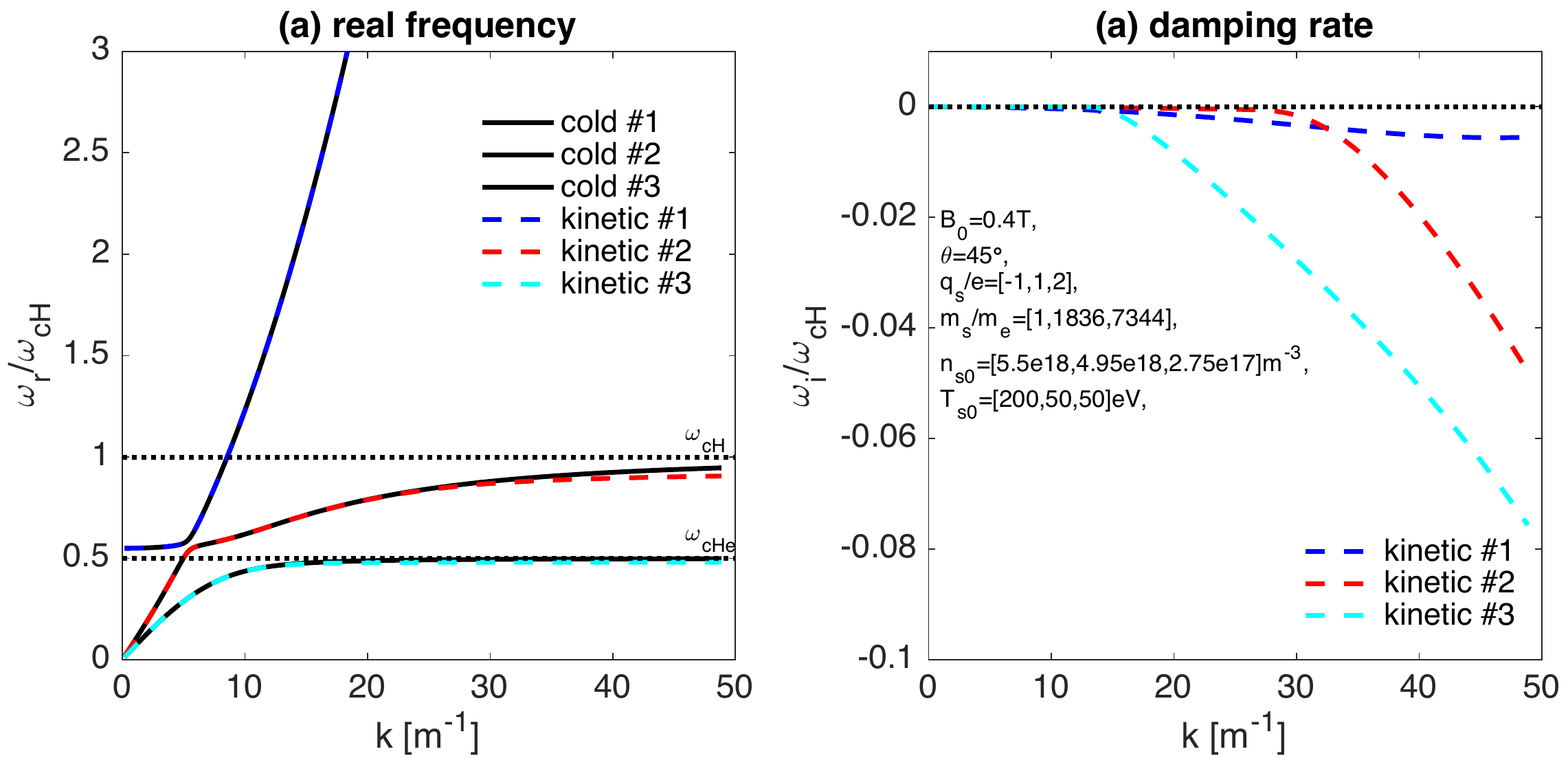}\\
\caption{Waves in ICW range for the EXL-50 He$^{2+}$ minority ions heating case. Solutions are calculated by the fluid and kinetic versions of BO\cite{Xie2019}.}\label{fig:EXL_BO_HHe_ICW}
\end{figure}

To analysis the wave feature in multi-species plasmas and to find the reasonable initial wave vector, the fluid and kinetic version of BO code can be useful, which can give all the wave frequency $\omega$ solutions for given wave vector $\bm k$ at one time without the requirement of initial guess frequency and thus will not miss solutions. We show a typical $\omega$ vs. $k$ figure in Fig.\ref{fig:EXL_BO_HHe_ICW} for ICRF minority heating parameter relevant to the case in Fig.\ref{fig:EXL_4_5MHz_HHe_ICW}. For this case, we can see that three branches exist in the ion cyclotron frequency range, and the kinetic correction to the cold plasma real frequency is small. Table \ref{tab:bofamily} summaries the role of each codes in BO family. 

\begin{table}[htp]
\caption{Fluid and kinetic plasma waves and instabilities code BO family\cite{Xie2014,Xie2016,Xie2019}.}
\begin{center}
{\scriptsize
\begin{tabular}{c|ccc}\hline
& Type  & Names & References\\\hline
& dispersion relation & PDRF,
PDRK, 
 & Xie14,
\\
BO family & ($\bm k\to\omega$) & BO,
BO2.0
 & 16,19,21
\\\cline{2-4}
(open source)& ray tracing & BORAY
 & Xie21
\\
&  ($\omega\to \bm k$)  & 
 & (this work)
\\\hline
\end{tabular}}
\end{center}
\label{tab:bofamily}

\end{table}%

\section{Analytical Solov\'ev equilibrium for varies configurations}\label{sec:solovev}
Analytical equilibrium can be useful for fast study the wave feature and can avoid the numerical interpolation error of numerical equilibrium from discrete grids. We construct an analytical Solov\'ev equilibrium to include tokamak, spherical tokamak, FRC and mirror configuration in a same model, and which is also the solution of Grad-Shafranov MHD equilibrium equation.

The normalized equilibrium poloidal flux is\cite{Helander2001} 
\begin{eqnarray}\label{eq:analyeq}
\psi(r,z)=-RA_\phi=\frac{\psi_0}{R_0^4}\Big\{(R^2-R_0^2)^2+\frac{Z^2}{E^2}(R^2-R_x^2)\\\nonumber
-\tau R_0^2\Big[R^2ln(\frac{R^2}{R_0^2})-(R^2-R_0^2)-\frac{(R^2-R_0^2)^2}{2R_0^2}\Big]\Big\},
\end{eqnarray}
where $R_0$ is major radius and the magnetic axis position $\psi(R_0,0)=0$. $R_x$, $E$ and $\tau$ control the position of $
X$-point, elongation and triangularity. The magnetic field are
\begin{eqnarray}\label{eq:}
B_r&=&-\frac{1}{r}\frac{\partial\psi}{\partial z}=-\frac{2\psi_0}{rR_0^2}\Big[\frac{Z}{E^2}(R^2-R_x^2)\Big],\\
B_z&=&\frac{1}{r}\frac{\partial\psi}{\partial r}=\frac{2\psi_0}{R_0^4}\Big\{2(R^2-R_0^2)+\frac{Z^2}{E^2}\\\nonumber
&&-\tau R_0^2\Big[ln(\frac{R^2}{R_0^2})-\frac{(R_x^2-R_0^2)}{R_0^2}\Big]\Big\}.
\end{eqnarray}
At X-point, $B_z(R_x,Z_x)=0$, which gives
\begin{eqnarray}\label{eq:}
Z_x=E\sqrt{\tau R_0^2ln(\frac{R^2}{R_0^2})+(2+\tau)(R_0^2-R_x^2)}.
\end{eqnarray}
Toroidal magnetic field  
\begin{eqnarray}\label{eq:}
B_\phi=\frac{B_0R_0}{R}.
\end{eqnarray}
Around magnetic axis ($z\to0$, $r\to R_0$), we can have
\begin{eqnarray}\label{eq:}
\psi=4\psi_0\epsilon^2,~\epsilon\equiv\frac{r-R_0}{R_0}\ll1,~\kappa\equiv\frac{2E}{\sqrt{1-R_x^2/R_0^2}}.
\end{eqnarray}
Thus poloidal magnetic field and safety factor around O-point is 
\begin{eqnarray}\label{eq:}
B_p=8\frac{\psi_0}{R_0^2}\epsilon,~~q_0=\frac{rB_0}{R_0B_p}=\frac{B_0R_0^2}{8\psi_0},
\end{eqnarray}
which gives
\begin{eqnarray}\label{eq:}
\psi_0=\frac{B_0R_0^2}{8q_0}.
\end{eqnarray}
The above model is very convenient to construct tokamak and spherical tokamak configurations. 

To construct FRC configuration, we set $\tau=0$, $R_x=0$ and  $B_\phi=0$, which yields Hill-vortex equilibrium. And we set the magnetic $B_z(0,0)=B_0$, which gives $\psi_0=\frac{B_0R_0^2}{4}$. The FRC model also holds for mirror configuration, we only need set further $R_0^2<0$.  That is, the Eq.(\ref{eq:analyeq}) can combine all the above several configurations in one model.

We construct the density and temperature profiles as
\begin{eqnarray}\label{eq:}
n_{s0}(r,z)=n_{s00}e^{-\frac{\psi}{\psi_xL_{ns}^2}},\\
T_{s0}(r,z)=T_{s00}e^{-\frac{\psi}{\psi_xL_{ts}^2}},
\end{eqnarray}
where $n_{s00}$ and $T_{s00}$ are density and temperature of species $s$ at O-point, and $L_{ns}$ and $L_{ts}$ are normalized scaling length of their radial profiles, with $\psi_x\equiv\psi(R_x,Z_x)$. Thus, the derivatives are readily obtained, say
\begin{eqnarray}\label{eq:}
\frac{\partial n_{s0}}{\partial r}=-\frac{1}{\psi_xL_{ns}^2}n_{s0}(r,z)\frac{\partial \psi}{\partial r}=-\frac{rB_z}{\psi_xL_{ns}^2}n_{s0},\\
\frac{\partial n_{s0}}{\partial z}=-\frac{1}{\psi_xL_{ns}^2}n_{s0}(r,z)\frac{\partial \psi}{\partial z}=\frac{rB_r}{\psi_xL_{ns}^2}n_{s0}.
\end{eqnarray}
The derivatives of magnetic field components are also readily obtained, and not shown here.

\end{document}